\documentstyle[pre,multicol,aps,epsfig]{revtex}
\begin{document}
\title{Multicomponent dipole-mode spatial solitons}

\author{Anton S. Desyatnikov and Yuri S. Kivshar}
\address{Nonlinear Physics Group, Research School of
Physical Sciences and Engineering, The Australian National University,
\\ Canberra ACT 0200, Australia}

\author{Kristian Motzek and Friedemann Kaiser}
\address{Institute of Applied Physics, Darmstadt University of
Technology, D-64289 Darmstadt, Germany}

\author{Carsten Weilnau and Cornelia Denz}
\address{Institute of Applied Physics, Westf\"alische
Wilhelms-Universit\"at M\"unster, D-48149 M\"unster, Germany}

\maketitle

\begin{abstract}
We study (2+1)-dimensional multicomponent spatial vector solitons
with a nontrivial topological structure  of their constituents,
and demonstrate that these solitary waves exhibit a
symmetry-breaking instability provided their total topological
charge is nonzero. We describe a novel type of stable
multicomponent dipole-mode solitons with intriguing swinging
dynamics.
\end{abstract}

\pacs{OCIS numbers: }

\vspace*{-0.9cm}

\begin{multicols}{2}
\narrowtext

Recent progress in the study of spatial optical solitons and their
interaction, as well as the extensive experimental demonstrations
of stable self-focussing of light in different types of nonlinear
bulk media,  open the road for new concepts to control the
diffraction of optical beams and to design new devices for optical
switching and storage~\cite{book}. Many novel fundamental concepts
recently suggested in the physics of spatial optical solitons are
associated with vectorial interaction and multicomponent soliton
beams that mutually self-trap in a nonlinear medium. Such
composite multimode solitons can have complex structures and, in
many cases, their total intensity profile exhibits multiple
humps~\cite{multi}.

In a bulk medium, vector solitons exist in different forms and, as
was recently shown for two-component self-trapped beams, many
types of multipole vector solitons can be predicted and analyzed
for an isotropic nonlinear bulk medium with saturable
nonlinearity~\cite{multi2}. Recently, an important generalization
of this concept to the case of $N-$component two-dimensional
vector solitons was suggested for an example of a thresholding
nonlinearity~\cite{OL}. In particular, Musslimani {\em et
al.}~\cite{OL} predicted the existence of multihump $N-$component
composite spatial solitons that carry different topological
charges (`spins') and, therefore,  can provide exciting
possibilities for `spin-dependant' interactions of self-trapped
optical beams \cite{inter}.

The purpose of this Letter is twofold. First, we study in more detail the dynamics of
multicomponent spatial solitons carrying topological charges in different components and
demonstrate that, in contrast to the conjecture of their stability made in Ref.~\cite{OL}, these
vector solitons demonstrate a symmetry-breaking instability in all the cases where their total
angular momentum is nonzero. Second, based on earlier studies of two-component vector
solitons~\cite{multi2} and the conceptual approach developed in Ref.~\cite{OL}, we propose a novel
type of stable multi-component vector solitons consisting of two perpendicular dipole components
trapped by the soliton-induced waveguide. These vector solitons are studied here for the case of
$N=3$ components, which are shown to be the building blocks for the solitons composed of $N$
incoherently coupled dipole-mode beams \cite{ourPRL}. Additionally, we demonstrate numerically
that these novel vector solitons are very robust for a broad range of their parameter space, and
they demonstrate intriguing swinging dynamics outside the stability domain, resembling long-lived
excitations and vibrations of molecules.

We consider the interaction of $N$ mutually incoherent (2+1)-dimensional optical beams propagating
in a bulk saturable medium, described by the normalized equations ($j=1, 2, \ldots, N$),
\begin{equation}
\label{eq1} i\frac{\partial E_j}{\partial z} +\Delta_\perp E_j - \frac{E_j}{1+ \Sigma |E_j|^2}=0,
\end{equation}
where $\Delta_\perp$ is the transverse Laplacian and $z$ is the propagation coordinate. Equations
(\ref{eq1}) describe, in a rather simplified isotropic approximation, screening spatial solitons
in photorefractive materials \cite{photo}.

To describe multicomponent vector solitons in the framework of the
model~(\ref{eq1}), first we look for stationary solutions in the
form $E_j(x,y,z)=u_j(x,y)\exp(-i\beta_j z)$, where $\beta_j$ is
the propagation constant and $u_j(x,y)$ is the envelope of the
$j$-th component. Then, introducing the dimensionless parameter
$\lambda_j=(1-\beta_j)/(1-\beta_1)$ and normalizing the field
amplitudes, $u_j\rightarrow\sqrt{1-\beta_1} u_j$, and the
coordinates, $(x,y)\rightarrow(x, y)/\sqrt{1-\beta_1}$, we obtain
\begin{equation}
\label{eq2} \Delta_\perp u_j - \lambda_j u_j + F(I)u_j=0,
\end{equation}
where $I=\Sigma |u_j|^2$ is the normalized total intensity, and $F(I)=I(1+sI)^{-1}$ with the
effective saturation parameter $s=1-\beta_1$.

First of all, following Musslimani {\em et al.}~\cite{OL}, we seek
multicomponent radially symmetric solutions of Eqs. (\ref{eq2})
for which the main component $u_1(x,y)=U_1(r)$ has no nodes, but
each of the components $u_k$ ($k>1$) carries a different
topological charge, $u_k(x,y)=U_k(r) \exp (im_k\theta)$. We denote
such states as ($0,..,m_k,..$), and an example for $N=3$ is
presented in Fig.~1(a), where the same intensity distribution
corresponds to {\em two different} states,  ($0,+1,+1$) and
($0,+1,-1$).

In order to study stability of these composite solitons, we
propagate them numerically and find that, provided the total
angular momentum is nonzero, all these multicomponent solitons
undergo a symmetry-breaking instability and fragment into a number
of the fundamental solitons, as shown in Fig. 1(b) for the case
$(0,+1,+1)$. This instability is similar to the instability of the
vortex-mode solitons described earlier for the two-component
model. The resulting incoherent superposition of two {\em
parallel} dipole components $u_2$ and $u_3$ can be regarded as a
generalization of a two-component {\em dipole-mode soliton} $\{
u_1,V\}$ \cite{dipole} to a three-component solution $\{ u_1 ,u_2
,u_3\}$ at $\lambda_2=\lambda_3$ with the help of a transformation
of the dipole components, $V\rightarrow \{ u_2,u_3 \}$, where $u_2
= V \cos\psi$ and $u_3 = V \sin\psi$ ($\psi$ is a transformation
parameter). Such a straightforward generalization is indeed
possible for the $N$-component system (\ref{eq2}).

\begin{figure}
\setlength{\epsfxsize}{7.5cm} \centerline{\epsfbox{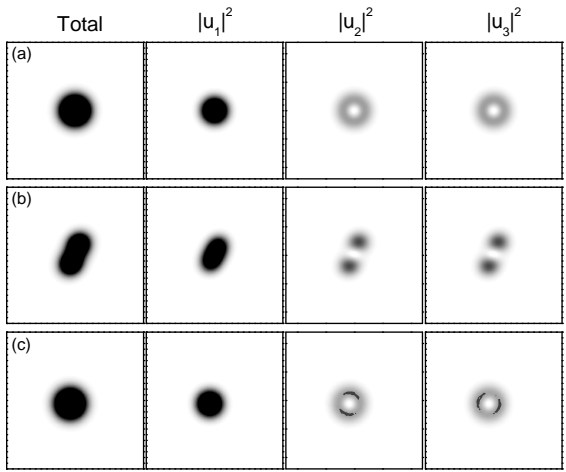}}
\vspace{1mm} \caption{ Evolution of the three-component soliton:
(a) stationary solution at $z=0$, (b) the symmetry-breaking
instability of the $(0,+1,+1)$ solution at $z=80$ , (c) the
long-lived quasi-stable propagation of the $(0,+1,-1)$ state at
$z=500$.}
\end{figure}

\vspace{-5mm}

\begin{figure}
\setlength{\epsfxsize}{7.5cm} \centerline{\epsfbox{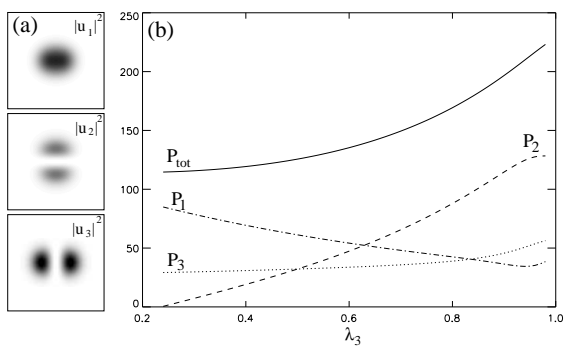}}
\vspace{1mm} \caption{ Families of the three-component dipole-mode
solitons. (a) Soliton structure at $\lambda_2=0.5$ and
$\lambda_3=0.65$, (b) the total and partial powers vs. $\lambda_3$
at fixed $\lambda_2 =0.5$.}
\end{figure}

\vspace{-5mm}

The most important property of the $(0,+1,-1)$ solution is that
its total angular momentum is {\em zero}, and this makes it
stable. In our calculations, this vector soliton was observed to
be {\em unchanged} for the distances of the order of $10^3$
diffraction lengths. However, being launched with additional
noise, this soliton displays {\em slowly growing modulations}, as
shown in Fig. 1(c). The total intensity of the modulated rings in
Fig. 1(c) preserves the initial ring profile, resembling an
incoherent superposition of two {\em perpendicular} dipole
components \cite{ourPRL}. While the vector soliton, consisting of
two crossed dipoles, has been shown to be {\em unstable} without
the third main component \cite{ourPRL}, we found that the
three-component dipole-mode soliton is stable in our numerical
simulations. Stabilization of the vector ring in the presence of
the third component can be explained by the physics of {\em the
soliton-induced waveguides}. Indeed, two crossed dipoles, $u_2$
and $u_3$, represent {\em a vectorial guided mode} of the induced
waveguide. A nontrivial rotational transformation of such a
solution (see Ref. \cite{ourPRL} for details) allows to find a
whole family of possible superpositions of these modes, including,
as a particular case, the vortex components shown in Fig. 1(a) and
the $N$-component dipole-mode soliton.

In order to find the multicomponent solitary waves with a
nontrivial geometry, we integrate the system (\ref{eq2})
numerically, by means of a relaxation technique, and find {\em a
novel class of the dipole-mode soliton} that consists of
perpendicularly oriented dipoles with {\em different powers}: the
simplest possible solution of this type has $N=3$ components, and
it is described by two independent parameters $(\lambda_2,
\lambda_3)$, as shown in Fig.~2(a). The family of these solitons
ranges from solutions where the fundamental mode dominates the
entire structure to solutions where one of the dipoles dominates,
as can be seen in Fig.~2(b), where for fixed $\lambda_2=0.5$ the
power of the components $P_j=\int\left|u_j\right|^2 d{\mathbf r}$
is shown as a function of $\lambda_3$.

Numerical propagation of these solitons has shown that from the
lower cutoff value for $\lambda_3$, where the intensity of the
$u_3$-component vanishes, up to a value of about $\lambda_3=0.7$
these vector solitons are stable, whereas for higher $\lambda_3$
they decompose to form different new structures. As can be seen
from Fig. 3, an unstable soliton breaks the symmetry along both
symmetry axes of the initial distribution. The products of this
instability (see the last row in Fig.~3) are a fundamental vector
soliton and a rotating dipole-mode soliton, recently introduced in
Ref.~\cite{prop} as {\em `a propeller soliton'}. Those two simpler
solitons fly away from each other after the break-up.

Near the instability threshold, for $0.7 < \lambda_3 < 0.8$, we
observe very interesting and intriguing dynamics, associated with
weak oscillatory instabilities. Figure 4 shows a characteristic
example of this dynamics, when the instability breaks the symmetry
only along one of the symmetry axes (parallel to the orientation
of the stronger dipole). The product of this instability is a
structure consisting of a tripole, a dipole and a nodeless beam.
This structure is remarkably long-lived and it has, as the
snapshots show, {\em a swinging behavior} resembling a swinging
mode of a three-atom molecule. We could observe almost three
periods of such oscillations, until a strong energy-exchange
between the two dipole beamlets sets in and destroys this
structure. We expect that the vibrational degrees of freedom,
which are likely associated with long-lived soliton internal
modes, should manifest themselves in the rich dynamics of soliton
collisions, as it is known for the study of a two-component
model~\cite{inter}.

\begin{figure}
\setlength{\epsfxsize}{7.5cm} \centerline{\epsfbox{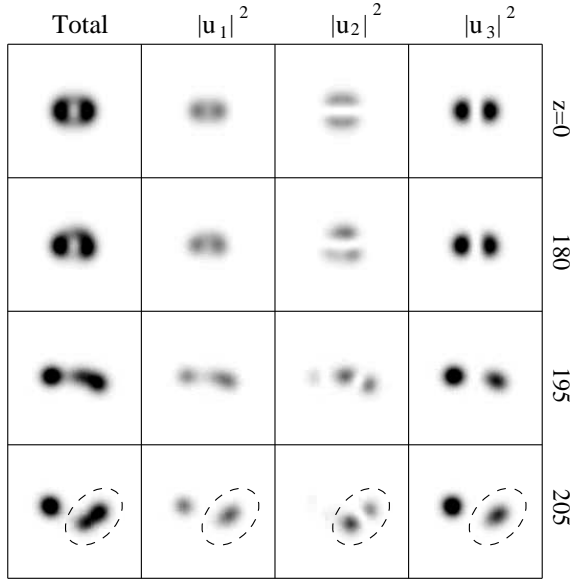}}
\vspace{1mm} \caption{ Unstable propagation of a three-component
dipole-mode soliton at $\lambda_2=0.5$ and $\lambda_3=0.8$, and
its decay into a fundamental vector soliton and a vector propeller
soliton (shown by dashed).}
\end{figure}

\vspace{-5mm}

\begin{figure}
\setlength{\epsfxsize}{7.5cm} \centerline{\epsfbox{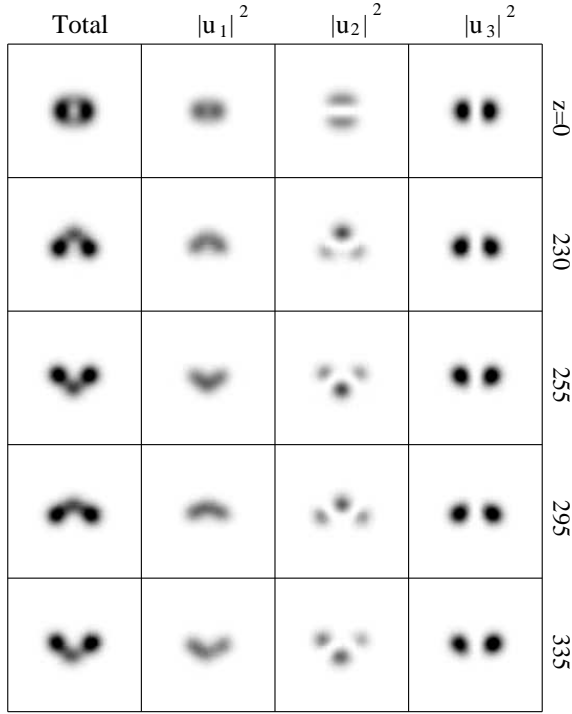}}
\vspace{1mm} \caption{ Swinging dynamics of the vector dipole-mode
soliton at $\lambda_2=0.5$ and $\lambda_3=0.75$.}
\end{figure}

\vspace{-3mm}

Having found these novel composite solitons for the isotropic
model, we wonder if the similar multicomponent solitons can exist
in an anisotropic nonlocal model which is more consistent with the
experimentally studied photorefractive
nonlinearities~\cite{aniso}. In order to verify this, we have used
the $N-$component generalization of the Zozulya-Anderson model
that takes into account the most important properties of
photorefractive nonlinearities~\cite{aniso2}, and found similar
classes of multicomponent localized solutions with perpendicularly
oriented dipole components. Since the anisotropy allows stable
stationary dipole modes oriented in two fixed directions
only~\cite{aniso}, these solutions are found to be stable even in
anisotropic media with nonlocal nonlinear response. This allows us
to expect the subsequent experimental observation of the novel
type of vector solitons and swinging dynamics described above.

This research was supported by the Australian-German Joint
Research Cooperation Scheme, the German Academic Exchange Service
(DAAD), and the Australian Photonics Cooperative Research Centre.

\end{multicols}
\end{document}